\documentclass[a4paper,12pt]{article}      
\usepackage{graphicx}
\usepackage{amsmath}
\usepackage{amsfonts}

\begin{document}

%%%%%%%%%%%%%%%%%%%%%%%%%%%%%%%%%%%%%%%%%%%%%%%%%%%%%%%%%%%%%%%%%
%%%%%%%%%%%%%%       for article.sty {begin}        %%%%%%%%%%%%%
%%%%%%%%%%%%%%%%%%%%%%%%%%%%%%%%%%%%%%%%%%%%%%%%%%%%%%%%%%%%%%%%%
%____________________preprint number______________________
\begin{flushright}			%<--HEP number
       CHIBA-EP-118\\
       HEP-TH/0005175\\
       \today
%       \tiny{Compiled \today}
\end{flushright}
%\hrule
\begin{center}
%_____________________title_______________________________

\renewcommand{\thefootnote}{\fnsymbol{footnote}}

{
\Large \sffamily \bfseries%
  {%
    A Note on the Quadratic Divergence in Hybrid Regularization
  }
}

\vspace{18pt}
%_____________________author_____________________________
{\large \sffamily \bfseries {Koh-ichi Nittoh}}
      \footnote{E-mail: tea@cuphd.nd.chiba-u.ac.jp}
\\
{\small\it
Center for Frontier Electronics and Photonics, Chiba University,\\
1-33 Yayoi-cho, Inage-ku, Chiba 263-8522, Japan}\\
\end{center}\vspace{18pt}
%_____________________abstract_____________________________
\begin{abstract}
We consider the quadratic divergence of the Yang-Mills theory
when we use the hybrid regularization method
consisting of the higher covariant derivative terms
and the Pauli-Villars fields.
By the explicit calculation of the diagrams,
we show that
the higher derivative terms for the ghost fields
are necessary 
for the complete cancellation of the quadratic divergence.
\end{abstract}

%%%%%%%%%%%%%%%%%%%%%%%%%%%%%%%%%%%%%%%%%%%%%%%%%%%%%%%%%%%%%%%%%
% body of paper
%%%%%%%%%%%%%%%%%%%%%%%%%%%%%%%%%%%%%%%%%%%%%%%%%%%%%%%%%%%%%%%%%
\renewcommand{\thefootnote}{\arabic{footnote}}
\setcounter{footnote}{0}

\section{Introduction}

%\paragraph{necessity of invariant regularizations}
%
An invariant regularization scheme is necessary
for the treatment of the ultraviolet divergence
in quantum gauge theories.
The dimensional regularization is known as
the  most powerful and popular method,
but it is not available for the theory
preserving the symmetry which depends on the space-time dimension,
like chiral gauge theory or topological field theory.
In such a case,
the hybrid regularization based on the higher covariant derivative
(HCD) method~\cite{Slavnov72,Slavnov77,Faddeev-Slavnov91}
is expected to be useful.

%\paragraph{history}
%
The HCD method is a partial regularization itself
because the higher derivative terms are introduced in a covariant way;
HCD terms render the propagators less divergent
but the vertices more divergent.
It is easily shown
by the calculation of the superficial degree of divergence
that some diagrams at one-loop level are left unregularized.
%
%For this reason, 
We have to introduce an additional regularization scheme
to regularize the divergence from these diagrams.

The Pauli-Villars (PV) regularization is suitable
for the additional regularization
from the standpoint of the invariant method,
because the PV regulators are constructed
in a chiral invariantly~\cite{Frolov-Slavnov93,Frolov-Slavnov94}
or parity invariantly~\cite{Kimura94,Nittoh-Ebihara98MPLA}
by an infinite number of them.
Applying such PV fields to the hybrid regularization,
the method is expected to be an invariant regularization scheme
that is available for the theory containing higher symmetries
like the supersymmetric gauge theories.
It has not been verified, however,
whether such special PV fields actually regularize
the intrinsically divergent theory
in the framework of the hybrid regularization:
at least the Yang-Mills (YM) theory must be regularized properly.

%\paragraph{HCD regularization in now days}
%
In recent years, %however,
on the other hand,
it was pointed out that
the original Slavnov's hybrid regularization scheme
does not give the correct value of
the coefficient of the renormalization group (RG)
$\beta$-function
when the YM theory is regularized by this scheme%
~\cite{Martin-Ruiz95}.
Nevertheless,
this problem was overcome by the minor modification of the scheme%
~\cite{Asorey-Falceto95,Bakeyev-Slavnov96},
and it was confirmed
that the modification is proper at the one-loop level
by an explicit calculation~\cite{Pronin-Stepanyantz97PL}.

Since only the logarithmic divergence plays an important role
in the calculation of the RG functions,
the other divergence,
the quadratic divergence in the case of four dimensions,
has not been seriously considered.
If the regularization works properly,
the quadratic divergence ought to be canceled out.
But in this scheme, the cancellation is not trivial
because the HCD term contributes to the quadratic divergence
and then increases the complexity of the divergence~%
~\cite{Pronin-Stepanyantz97PL}.
So it is worthy to show the cancellation of the quadratic divergence
for the consistency of %this regularization scheme.
the hybrid regularization scheme.

%\paragraph{Our purpose of this note}
%
In this note, we confirm that
the quadratic divergence is actually canceled out in the YM theory
with the hybrid regularization of the HCD and %the infinite number of PV fields.
the PV method.
By an explicit calculation of the vacuum polarization tensor,
it is shown that
the higher derivative terms for the ghost fields are necessary for
the complete regularization of this method.
Using an infinitely many PV fields as the additional regulator,
we also check the consistency of such PV fields
in the quadratic divergence
when they are used in the hybrid regularization.

\section{The regularization method}

We consider the SU($N$) Yang-Mills theory
in four dimensional Euclidean space-time.
The action is given by
\begin{equation}
S=S_\mathrm{YM}+S_\mathrm{GF},
\label{eq:classical action}
\end{equation}
where 
\begin{align}
S_\mathrm{YM}&=
\frac{1}{ 4}\int \mathrm{d}^4x
F_{\mu\nu}^a F^{\mu\nu}{}^a,
\label{eq:YM action}\\
S_\mathrm{GF}&=
\int \mathrm{d}^4x
\left[
 \frac{\xi_0}{ 2}b^a b^a
 -b^a (\partial^\mu A_\mu^a)
 +\overline c ^a (\partial_\mu D^\mu c)^a
\right],
\label{eq:GF action for YM}
\end{align}
with the field strength 
$
F_{\mu\nu}^a = 
\partial_\mu A_\nu^a - \partial_\nu A_\mu^a
+g f^{abc}A_\mu^b A_\nu^c
$
and the covariant derivative
$
D_\mu^{ac} =
\delta^{ac}\partial_\mu
+ gf^{abc}A_\mu^b.
$
Here $A_\mu^a$, $c^a$, $\overline c ^a$ and $b^a$
denote the gauge field, ghost, anti-ghost
and auxiliary field respectively,
$\xi_0$ is the gauge-fixing parameter 
and $f^{abc}$ is the structure constant of the gauge group.

%\paragraph{Steps}
The hybrid regularization consists of the following two steps:
first we introduce HCD terms and next PV fields.
The HCD terms improve the behavior of propagators at large momentum,
rendering the theory less divergent
at the cost of the emergence of new vertices,
and the theory is reduced to superrenormalizable,
i.e. there are just a finite number of divergent loops.
As see later, 
all the diagrams except one-, two-, three- and four-point functions
at one-loop level are convergent
with a suitable choice of the HCD action.
We deal with the remaining divergence by a PV type of regularization.

\subsection{Introduction of HCD terms}

We first regularize the action by an addition of the HCD term;
\begin{equation}
S_\Lambda = S_\mathrm{YM}+S_\mathrm{HCD}+S_\mathrm{GF}^H.
\label{eq:regularized action}
\end{equation}
$S_\mathrm{HCD}$ is the HCD action
and $S_\mathrm{GF}^H$ is the modified gauge-fixing action
when we use the HCD method.
The explicit forms are 
\begin{align}
S_\mathrm{HCD} &=
\frac{1}{ 4 \Lambda^4}\int \mathrm{d}^4x
(D^2F_{\mu\nu})^a (D^2F^{\mu\nu})^a,
\label{eq:HCD action} \\
S_\mathrm{GF}^H &=
\int \mathrm{d}^4x
\left[
 \frac{\xi_0 }{ 2}
 b^a b^a
 -b^a H(\partial^2/\Lambda^2) (\partial^\mu A_\mu^a)
 +\overline c ^a H(\partial^2/\Lambda^2) (\partial_\mu D^\mu c)^a
\right],
\label{eq:GF with HD action}
\end{align}
where $\Lambda$ is a cutoff parameter which has mass dimension of one,
and $H(\partial^2/\Lambda^2)$ 
is a dimensionless function
and must be a polynomial of $\partial^2/\Lambda^2$
to ensure the locality of the gauge field.
Its explicit form is determined
by the behavior of the gauge propagator
which is obtained
from (\ref{eq:regularized action}) as follows:
\begin{equation}
\frac{\Lambda^4 }{ p^4 (p^4 + \Lambda^4)}
(p^2 \delta_{\mu\nu} - p_\mu p_\nu)
+
\frac{\xi_0 }{ p^4 H^2({p^2 }/{ \Lambda^2})}
p_\mu p_\nu.
\label{eq:gauge propagator}
\end{equation}
The first term has the momentum degree of $-6$,
so the second term must be the same degree or less
to ensure the convergence of the diagrams;
%except one-, two-, three- and four-point functions at one-loop level;
%
$H^2$ behaves $\sim p^4$ or higher at large $p$.
While
the familiar form of the propagator must be recovered
in the limit of $\Lambda \rightarrow \infty$;
$H^2$ converges to unity at large $\Lambda$.
On these conditions,
the simplest choice of the $H^2$ in momentum space is%
\begin{equation}
H^2
\left(\frac{p^2 }{ \Lambda^2}\right)
=
1 + \frac{p^4 }{ \Lambda^4}.
\label{eq:explicit H^2}
\end{equation}
Since the first and the second term of \eqref{eq:gauge propagator}
has the same denominator in this choice of $H^2$,
when we work in the Feynman gauge ($\xi_0=1$),
the gauge propagator is reduced to
\begin{equation}
\frac{\Lambda^4 }{ p^2 (p^4 + \Lambda^4)}
\delta_{\mu\nu}.
\end{equation}
We take the Feynman gauge in the explicit calculation of diagrams
in the following section.

The HCD action is generally introduced in the form of
$
  \frac{1}{ 4 \Lambda^{2n}}\int \mathrm{d}^4x
  (D^nF_{\mu\nu})^2% (D^nF^{\mu\nu})^a
$.
In such a case,
the superficial degree of divergence is written 
\begin{equation}
\omega = 4 - 2n (L-1)
- E_A
%- (n+1)E_c,
\label{eq:degree of divergence}
\end{equation}
where $L$ and $E_A$ are the number of loops and
the number of the external lines of the gauge field.
For all the diagrams higher than two-loops ($L\ge2$),
$n \ge 2$ always gives negative $\omega$.
This means that
we may remove the higher loops by a suitable choice of $n$.
%
%\subparagraph{$L=1$}
%
For one-loop ($L=1$),
$\omega$ is not always negative for any $n$.
% %
As we take the higher $n$,
though the propagator becomes more convergent at large momentum,
the vertices are more divergent and increase their complexities.
Then the calculation of the diagram is more complex
even though at one-loop level.
The most economical choice is $n=2$
which leads the HCD action~\eqref{eq:HCD action}.

\subsection{Introduction of Pauli-Villars fields}

So far, all the diagrams except one-, two-, three- and four-point
functions at one-loop level are convergent by the HCD action.
We deal with the remaining divergence by a PV type of regularization.

Consider the following generating functional:
\begin{multline}
Z[J,\chi,\eta,\overline\eta]=
   \int \mathcal{D}A_\mu \mathcal{D}b
        \mathcal{D}\overline c \mathcal{D}c
\,\exp[-S_\Lambda - S_J]\\
\prod_{j=1}^\infty
 \Big[\det{}^{-\frac{\alpha_j}{2}}\mathbf{A}_j\Big]
%\prod_{j=1}^\infty 
 \Big[\det{}^{-\frac{\alpha_{-j}}{2}}\mathbf{A}_{-j}\Big]
\prod_{i=1}^\infty 
 \Big[\det{}^{\gamma_i}\mathbf{C}_i\Big]
%\prod_{i=1}^\infty
 \Big[\det{}^{\gamma_{-i}}\mathbf{C}_{-i}\Big],
\label{eq:generating functional}
\end{multline}
where $S_J$ is a source term consisting from
$J$, $\chi$, $\eta$ and $\overline \eta$
which is the source of $A_\mu$, $b$, $\overline c$ and $c$
respectively.
%
%\subparagraph{PV fields}
$\Big[\det{}^{-\frac{\alpha_j}{2}}\mathbf{A}_j\Big]$ and
$\Big[\det{}^{\gamma_i}\mathbf{C}_i\Big]$ are PV determinants
for the gauge and ghost field respectively,
defined by
\begin{align}
\Big[\det{}^{-\frac{\alpha_j}{2}}\mathbf{A}_j\Big]&=
\int \mathcal{D}A_j{}_\mu \mathcal{D}b_j
\exp[-S_{M_j}-S_{b_j}],
\label{eq:det A} \\
\Big[\det{}^{\gamma_i}\mathbf{C}_i\Big]&=
\int \mathcal{D}\overline c_i\mathcal{D}c_i
\exp[-S_{m_i}],
\label{eq:det C}
\end{align}
where
$\{\alpha_j\}$ and $\{\gamma_i\}$ are real parameters
to be fixed for each indices 
and the explicit forms of the functions in the exponents are
\begin{align}
S_{M_j}&=
\frac{1}{ 2}\int \mathrm{d}^4x \mathrm{d}^4y \nonumber \\
&\qquad \qquad
A_j{}_\mu^a(x)
\left[
 \frac{\delta^2 S_\Lambda }{ \delta A_\mu^a(x) \delta A_\nu^b(y)}
 -M_j^2 \delta^{ab} g^{\mu\nu} \delta(x-y)
\right]
A_j{}_\nu^b(y),
\label{eq:PV for gauge action} \\
S_{b_j}&=
\int \mathrm{d}^4x
\left[
 \frac{\xi_j }{ 2}
 b_j^a b_j^a
 -b_j^a {\Tilde H}(D^\mu A_j{}_\mu^a)
\right],
\label{eq:auxiliary field for PV action}\\
S_{m_i}&=
\int \mathrm{d}^4x
\left[
 \overline c_i^a {\Tilde H}(D_\mu D^\mu c_i)^a
 -m_i^2 \overline c_i^a c_i^a
\right].
\label{eq:PV for ghost action}
\end{align}
The field $A_j{}_\mu^a$ is a PV field for the gauge of
mass $M_j$,
$b_j^a$ an auxiliary field for $A_j{}_\mu^a$,
$\overline c_i^a$ and $c_i^a$ 
PV fields for the ghost and anti-ghost of mass $m_i$.
We introduce a gauge-fixing parameter $\xi_j$
for the correct regularization of the theory%
~\cite{Asorey-Falceto95}.
${\Tilde H}$ is the HCD term for the `gauge-fixing function'
for the PV fields,
which has the form
\begin{equation}
{\Tilde H}=\left(1+\frac{D^4}{\Lambda^4}\right)^\frac{1}{2}.
\end{equation}

These PV fields have the same form as the ones
that we used in the Chern-Simons gauge theory%
~\cite{Nittoh-Ebihara98MPLA,Nittoh99phd}.
The idea %of this method 
is to regularize the theory
with the pairs of the two types of the PV fields
$A_j$ and $A_{-j}$.
In the Chern-Simons gauge theory
to construct the parity invariant regulator,
the two fields are
related by the parity transformation
and represented by the slightly different actions,
but in this case
they have the same action except the sign of the index.

%%%%{the reason for infinite}
%
The reason why we have to introduce an infinite number of the PV fields
comes from the idea `to regularize with the pair'.
Imagine when the gauge field is regularized
by the PV pairs which we prepared above.
Since the number of the gauge field is one,
introducing one pair corresponds to subtracting double the divergence.
Then to remedy the over subtraction
we introduce another pair of opposite statistics
which means adding double the divergence.
To remedy the over addition
we have to introduce the third pair.
Such steps correspond to introducing
fermionic PV fields $(\alpha_j=-1)$
and bosonic PV fields $(\alpha_j=+1)$ alternately.
We repeat such steps alternately until the divergence is removed.
Namely,
we cannot regulate the theory by a finite number of PV pairs,
but we need an infinite number.
Then we take the following as PV conditions:
\begin{align}
M_j&=M|j|,&
\alpha_j&=(-1)^j.
\label{eq:PV conditions}
\end{align}
In the same way,
we take the PV conditions for ghost and anti-ghost
such as $m_i=m|i|$ and $\gamma_i=(-1)^i$.

%\paragraph{BRST invariance of the action}

The generating functional \eqref{eq:generating functional}
is invariant under the BRST transformations
in the reference~\cite{Nittoh-Ebihara98MPLA},
%the regulators are introduced BRST invariantly.
and this regularization manifestly preserves BRST invariance.

\subsection{Feynman rules}

The regularized action $S_\Lambda$ is decomposed
into the kinetic part $K$ and the vertex part $V$ as follows%
~\cite{Pronin-Stepanyantz97PL,Pronin-Stepanyantz97NP}:
\begin{equation}
\int \mathrm{d}^4x \Psi (x) (K + V + M^2) \Phi (x),
\end{equation}
where $\Psi(x)$ and $\Phi(x)$ denotes
an arbitrary field and $M$ its mass term.
Since $K$ and $V$ consist of the $\Lambda$-free
part from the YM term
(we denote with suffix `0')
and the $\Lambda$-dependent part from HCD term
(with suffix `$\Lambda$')
we formally decompose 
\begin{align}
K&=K_0+\frac{1}{\Lambda^4}K_\Lambda,&
V&=V_0+\frac{1}{\Lambda^4}V_\Lambda.
\end{align}
Then the propagators are written in the form
\begin{equation}
\frac{1}{K+M^2}
=\frac{1}{K_0+M^2}
 \left(
   1-\frac{K_\Lambda}{K_0+M^2}\Lambda^{-4}+O(\Lambda^{-8})
 \right).
\end{equation}
Using this decomposition,
the Feynman rules are written by the order of $\Lambda$
and we calculate the quantum corrections
%in its orders.
order by order of $\Lambda$.

\section{One-Loop Contributions}

Now we calculate the one-loop vacuum polarization tensor
order by order in $\Lambda$ up to $\Lambda^{-4}$.
There are eleven diagrams in $\Lambda^0$ order
and twelve diagrams in $\Lambda^{-4}$ order.
Each diagram has the quadratic divergence
which must cancel in totally.
We consider this divergence calculating
each diagram under the Feynman gauge $\xi_j=1$.
The calculation is carried out under the same rules 
that we take in references%
~\cite{Nittoh-Ebihara98MPLA,Nittoh99phd,NittohEP122}
whose summaries are the following:

\begin{enumerate}

\item
Take the same assignment for the internal momentum
among the graphically same diagrams.

\item
Take the infinite sum of the graphically same diagrams
under the PV condition \eqref{eq:PV conditions}.

\item
If there is no massless term
(whose index $j$ or $i$ is zero)
for the infinite sum,
add the lacking terms and subtract the same ones to balance.

\end{enumerate}

All the diagrams are classified into three groups
by the kind of the internal line in the diagram
and calculation is carried out in each group.

First we consider the diagrams
that contains only $A_j{}_\mu$ fields in the internal lines.
The index $j$ runs from $-\infty$ to $\infty$
except zero for these diagrams,
if we take the infinite sum from $-\infty$ to $\infty$
we have to add the diagrams which contains `$A_0{}_\mu$' field
in the internal lines.
Fortunately,
the diagrams in which the gauge field runs
have the same structure of them
and we take the infinite sum from $-\infty$ to $\infty$
without any extra terms.
Then the total of the quadratic divergence from these diagrams
is calculated
\begin{equation}
\frac{g^2 c_v \delta^{ab}}{8\pi^2}
\left(
 \frac{M^2}{40}C_2 - \frac{9M^6}{154\Lambda^4}C_4
\right)\delta_{\mu\nu}.
\label{eq:quadratic divergence related to the gauge}
\end{equation}
Where $C_2$ and $C_4$ are the constants
arising from the infinite sum of the index $j$%
~\cite{Nittoh99phd,NittohEP122}.

For the diagrams that contains the $b_j$ fields,
there is no diagrams from the field $b$
to compensate the diagrams consisting of $b_0$
because any vertex of the form $\left< A_\mu A_\nu b \right>$
does not exist in the theory.
So we have to add the $b_0$ diagrams to take the infinite sum
and subtract the same after the summations to balance out.
The total contributions are calculated
\begin{multline}
-\frac{g^2 c_v \delta^{ab}}{8\pi^2}
\left(
 \frac{M^2}{40}C_2 - \frac{9M^6}{154\Lambda^4}C_4
\right)\delta_{\mu\nu}
\\
+g^2c_v\delta^{ab}
 \int \frac{\mathrm{d}^4 k}{(2\pi)^4}\frac{1}{k^2(k-p)^2}
  \left[
     2k^2\delta_{\mu\nu}-3k_\mu k_\nu
    +\frac{1}{\Lambda^4}
      \left( 2k^6\delta_{\mu\nu} - 4k^4 k_\mu k_\nu \right)
  \right].
\label{eq:quadratic divergence related to b_j}
\end{multline}
The first line comes from the infinite sum with index $j$
and the second line is the counter terms
which we introduced to take the sum.

The situation does not alter in the diagrams
containing $\overline c_i$ and $c_i$ in the internal lines.
Since there are some differences in the vertices
between ghost fields and PV fields for ghosts,
$\overline c$ and $c$
do not play the role of $\overline c_0$ and $c_0$.
Then we have to add some massless terms for the infinite sums
and subtract the same ones
in the same way as the diagrams with $b_j$.
We calculate the contributions from these diagrams
as follows:
\begin{equation}
-g^2c_v\delta^{ab}
 \int \frac{\mathrm{d}^4 k}{(2\pi)^4}\frac{1}{k^2(k-p)^2}
  \left[
     2k^2\delta_{\mu\nu}-3k_\mu k_\nu
    +\frac{1}{\Lambda^4}
      \left( 2k^6\delta_{\mu\nu} - 4k^4 k_\mu k_\nu \right)
  \right].
\label{eq:quadratic divergence related to the ghosts}
\end{equation}
The each contribution has the quadratic divergence
proportional to the constant $C_2$ and $C_4$
after the infinite sum with indices,
but these divergence cancel out in total
within this group.
Then only the quadratic divergence from the counter terms remain
as in \eqref{eq:quadratic divergence related to the ghosts}.

It is easy to see from
\eqref{eq:quadratic divergence related to the gauge},
\eqref{eq:quadratic divergence related to b_j} and
\eqref{eq:quadratic divergence related to the ghosts}
that the quadratic divergence of the vacuum polarization tensor
disappears from the theory.

%\paragraph{when used $f$}
%
Here we notice that
the function $f(\partial^2/\Lambda^2)$ in the reference~\cite{Martin-Ruiz95}
does not cancels the quadratic divergence completely.
In that reference,
since the modified gauge-fixing action is inserted through the form
$\frac{\xi_0}{2f}b^2-b\partial^\mu A_\mu$
instead of \eqref{eq:GF with HD action},
so the naively extended gauge-fixing action for the PV field is written by
$\frac{\xi_j}{2f}b_j^2-b_j\partial^\mu A_j{}_\mu$.
This action, however, breaks the BRST invariance
because $f$ contains usual derivative $\partial_\mu$.% by definition.
This symmetry breaking effects to the cancellation of the quadratic divergence.
In such a case,
all the Feynman rules 
related to the auxiliary, ghost and their PV fields
are modified in $\Lambda^{-4}$ order,
and then the quadratic divergence corresponds to
\eqref{eq:quadratic divergence related to b_j} and
\eqref{eq:quadratic divergence related to the ghosts}
are calculated as follows:
\begin{multline}
-\frac{g^2c_v\delta^{ab}}{8\pi^2}
\left(
 \frac{M^2}{40}C_2 - \frac{6M^6}{154\Lambda^4}C_4
\right)\delta_{\mu\nu}\\
+g^2c_v\delta^{ab}
 \int \frac{\mathrm{d}^4 k}{(2\pi)^4}\frac{1}{k^2(k-p)^2}
  \left[
     2k^2\delta_{\mu\nu}-3k_\mu k_\nu
  \right],
\tag{\ref{eq:quadratic divergence related to b_j}$^\prime$}
\end{multline}
\begin{equation}
-g^2c_v\delta^{ab}
 \int \frac{\mathrm{d}^4 k}{(2\pi)^4}\frac{1}{k^2(k-p)^2}
  \left[
     2k^2\delta_{\mu\nu}-3k_\mu k_\nu
  \right].
\tag{\ref{eq:quadratic divergence related to the ghosts}$^\prime$}
\end{equation}
This result shows that
the complete cancellation at $\Lambda^{-4}$ does not occur
though the contribution at $\Lambda^0$ cancels.
We may choose the higher \textit{covariant} derivative function $\Tilde f$
instead of $f$ to avoid this difficulty
as we extended $H$ to $\Tilde H$.
If we substitute $\Tilde f = 1+\frac{D^4}{\Lambda^4}$
in place of $f = 1+\frac{\partial^4}{\Lambda^4}$,
some new diagrams arise from the new vertices among $b_j$
under the naively treatment of $\Tilde f$.
These diagrams, however, give no quadratic divergence to
(\ref{eq:quadratic divergence related to the ghosts}$^\prime$) in total,
and then the divergence is not removed.
This failure may be caused by the treatment of the non-local contribution
in the function $\Tilde f$.
We may have to consider the exact treatment of such terms
when we start from the action described by $f$.

\section{Conclusion}

In this note
we consider the cancellation of the quadratic divergence
of the YM theory
regularized by the hybrid regularization
consisting of the HCD method and the infinitely many PV fields.
By an explicit calculation of the vacuum polarization tensor
up to $\Lambda^{-4}$ order,
all the quadratic divergence exactly cancels in all orders.
This result shows that
the quadratic divergence is regularized by the hybrid regularization
as well as the logarithmic divergence.
%

%\paragraph{cancellation mechanism}
%The cancellation occurs
The divergence cancels
order by order in $\Lambda^{-4}$
and the cancellation mechanism is the same in all orders:
the combination of
\eqref{eq:quadratic divergence related to the gauge} and
\eqref{eq:quadratic divergence related to the ghosts}
cancels with 
\eqref{eq:quadratic divergence related to b_j}.
We expect that
this mechanism works
in all the higher orders than $\Lambda^{-4}$
e.g. in the order of $\Lambda^{-8}$
and the quadratic divergence completely cancels out.

%\paragraph{necessary of HCD for ghost}
In our calculation,
the higher derivative term for the ghost field, $H$,
plays an important role in the cancellation of the quadratic divergence.
Since any contribution of $H$ does not appear
in the superficial degree of divergence $\omega$
in \eqref{eq:degree of divergence},
the necessity of $H$ is unclear.
So the simplest choice of the higher derivative term
to improve the longitudinal part of the gauge propagator
is the function $f$ in the reference~\cite{Martin-Ruiz95}.
In such a case,
the treatment of the non-local contribution is so problematic
that the cancellation of the quadratic divergence
is not shown by the calculation.
$H$, however, is introduced instead of $f$ at the beginning,
such a difficulty does not arise
and we can demonstrated the cancellation clearly.

%\paragraph{logarithmic divergence}
%
We are also interested in the coefficient of
the $\beta$-function with this regularization scheme.
Since $H$ gives the same effect with $f$ 
and does not give any contribution to the logarithmic divergence
in the order of $\Lambda^0$
we get the familiar value of the coefficient.
We will discuss in detail
the logarithmic divergence of this theory elsewhere%
~\cite{NittohEP122}.

%_______________________ References ________________________
%
%
% \bibliographystyle{prsty}
% %\bibliographystyle{plain}
% %\addcontentsline{toc}{chapter}{Bibliography}
% \bibliography{quad_div}

\begin{thebibliography}{10}

\bibitem{Slavnov72}
A. Slavnov, Theor. Math. Phys. {\bf 13},  1064  (1972).

\bibitem{Slavnov77}
A. Slavnov, Theor. Math. Phys. {\bf 33},  210  (1977).

\bibitem{Faddeev-Slavnov91}
L. Faddeev and A. Slavnov, {\em {Gauge Field, Introduction to Quantum Theory}},
  2nd ed. (Addison-Wesley, Redwood, 1991).

\bibitem{Frolov-Slavnov93}
S. Frolov and A. Slavnov, Phys. Lett. {\bf B309},  344  (1993).

\bibitem{Frolov-Slavnov94}
S. Frolov and A. Slavnov, Nucl. Phys. {\bf B411},  647  (1994).

\bibitem{Kimura94}
T. Kimura, Prog. Theor. Phys. {\bf 92},  693  (1994).

\bibitem{Nittoh-Ebihara98MPLA}
K. Nittoh and T. Ebihara, Mod. Phys. Lett. {\bf A13},  2231  (1998).

\bibitem{Martin-Ruiz95}
C. Martin and F. Ruiz~Ruiz, Nucl. Phys. {\bf B436},  545  (1995).

\bibitem{Asorey-Falceto95}
M. Asorey and F. Falceto, Phys. Rev. {\bf D54},  5290  (1996).

\bibitem{Bakeyev-Slavnov96}
T. Bakeyev and A. Slavnov, Mod. Phys. Lett. {\bf A11},  1539  (1996).

\bibitem{Pronin-Stepanyantz97PL}
P. Pronin and K. Stepanyants, Phys. Lett. {\bf B414},  117  (1997).

\bibitem{Nittoh99phd}
K. Nittoh, {Ph.D.} dissertation, Chiba University, 1999.

\bibitem{Pronin-Stepanyantz97NP}
P. Pronin and K. Stepanyants, Nucl. Phys. {\bf B485},  517  (1997).

\bibitem{NittohEP122}
K. Nittoh, Chiba Univ. preprint, CHIBA-EP-122 (in preparation).

\end{thebibliography}
% %...........................................................
%___________________________________________________________

\end{document}